\documentclass{iau-JDSS}
\usepackage{graphicx}

\title[A Zoo of Galaxies] 
{Invited Discourse: ``A Zoo of Galaxies"}

\author[Karen L. Masters]   
{Karen L. Masters}

\affiliation{$^1$Institute for Cosmology and Gravitation, University of Portsmouth, Dennis Sciama Building, Burnaby Road, Portsmouth, PO1 3FX, UK \break email: karen.masters@port.ac.uk \\[\affilskip] $^2$South East Physics Network, www.sepnet.ac.uk}

\pubyear{2013}
\volume{Volume 16}  
\pagerange{}
\date{?? and in revised form ??}
\jname{Highlights of Astronomy, Volume 16}
\editors{Thierry Montmerle, ed.}

\begin{document}

\maketitle

\begin{abstract}
We live in a universe filled with galaxies with an amazing variety of sizes and shapes. One of the biggest challenges for astronomers working in this field is to understand how all these types relate to each other in the background of an expanding universe. Modern astronomical surveys (like the Sloan Digital Sky Survey) have revolutionised this field of astronomy, by providing vast numbers of galaxies to study. The sheer size of the these databases made traditional visual classification of the types galaxies impossible and in 2007 inspired the Galaxy Zoo project (www.galaxyzoo.org); starting the largest ever scientific collaboration by asking members of the public to help classify galaxies by type and shape. Galaxy Zoo has since shown itself, in a series of now more than 30 scientific papers, to be a fantastic database for the study of galaxy evolution. In this Invited Discourse I spoke a little about the historical background of our understanding of what galaxies are, of galaxy classification, about our modern view of galaxies in the era of large surveys. I finish with showcasing some of the contributions galaxy classifications from the Galaxy Zoo project are making to our understanding of galaxy evolution.
\end{abstract}

\firstsection 
\section{What are Galaxies?}

\begin{figure}
\includegraphics[width=4in]{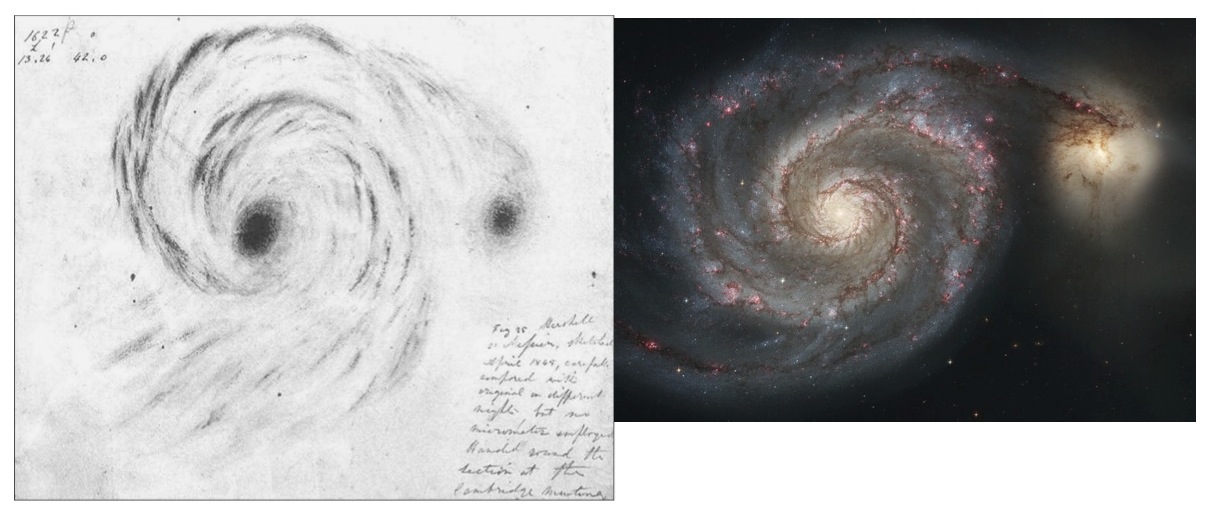}
  \caption{Two views of M51 (the Whirlpool Galaxy). On the left is the original 1845 drawing by William Parsons (Figure 8 from Steinick 2012). On the right is an image from the Hubble Space Telescope (Credit: NASA, ESA, S. Beckwitch (STSci) and the Hubble Heritage Team (STSci/AURA)).    \label{fig:M51}}
\end{figure}

To say our understanding of the ``zoo" of galaxies that are found in our Universe has changed a lot over the last century or two is a bit of an understatement. In 1845 the state of the art picture of an external galaxy, was an image of M51, or the Whirlpool galaxy drawn by William Parsons, Third Earl of Rosse (1800-1867), looking through what at the time was the largest telescope in the world - the ``Leviathon of Parsontown" at his castle in Ireland (Left panel Figure \ref{fig:M51}; for a discussion of the history of picture see Steinicke 2012),  At the time no-one fully understood that this was an external galaxy made of billions of stars. Although one of the motivations Parson is said to have had to build the telescope, was to resolve structure in distant nebula to see if they could be ``island universe", rather similar in structure to our own galactic home. Today we have images of millions of galaxies from large surveys like the Sloan Digital Sky Survey (e.g. the last imaging release in Data Release Eight, Aihara et al. 2011), and extraordinarily high resolution images of hundreds of galaxies, including the Whirlpool, from the Hubble Space Telescope (Fig \ref{fig:M51}, right), as well as information on galaxies extending back more than half of the age of our Universe. 

\begin{figure}
\includegraphics[width=3in]{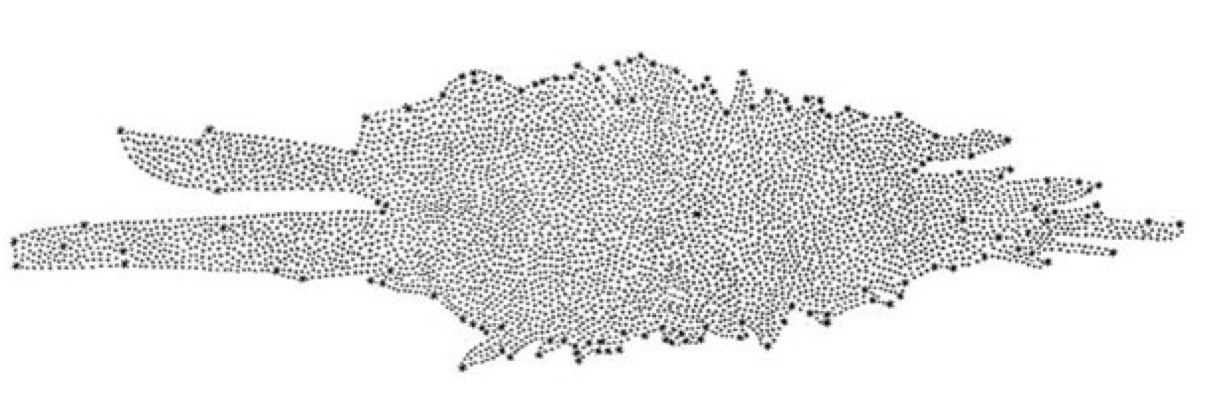}
  \caption{The first map of the Milky Way showing it as a collection of stars, and based on observational data. (Herschel 1785)   \label{fig:HerschelMW}}
\end{figure}

  It took a long time for astronomers to understand that stars in the universe are organized in collections we now call galaxies. I can't help thinking that it was a major leap in understanding for astronomers to connect the uneven distribution of stars in the night sky with the three-dimensional structure of the galaxy that we live in. The first published example of this idea is the map shown in Figure \ref{fig:HerschelMW}, published by William Herschel in 1785 (Herschel 1785), and based on star counts made by himself and his sister Caroline. This diagram demonstrates an understanding of the Galaxy as a collection of stars, and while there is a a lot wrong with it (for example the Sun is at the centre, and the whole thing was much too small at $\sim7000$ light years across) itÕs  an extraordinary piece of work as the first example of such a map based on actual astronomical data. 

\begin{figure}
\includegraphics[height=2.5in,angle=0]{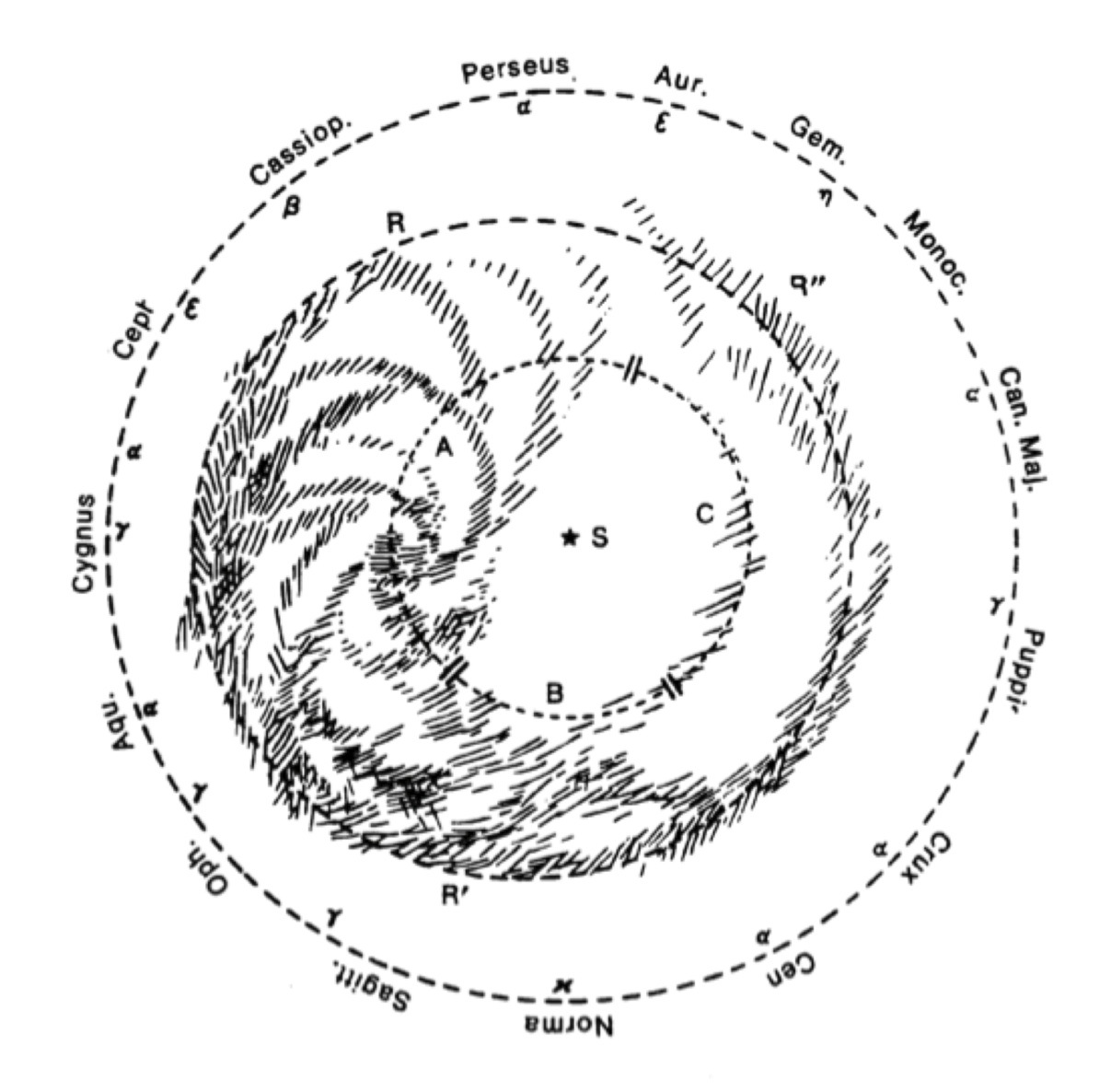}
  \caption{Cornelius Easton's model of the Galaxy in 1900 (reproduced from Fig 2 of Trimble 1995) \label{fig:easton}}
\end{figure}

By 1900 astronomers understood quite a lot more about the basic structure of our galaxy. 
The map made by Cornelius Easton in 1900 (Fig \ref{fig:easton}) was the first to show our galaxy as having spiral structures (Trimble 1995).  Easton used pictures of other spiral galaxies he saw in the sky to suggest the Milky Way might have this structure, although he still incorrectly placed the Sun in the centre of the Galaxy, and it's still too small.

 The variety of different galaxies observed in the sky naturally caused people to wonder what they were. The scientific arguments surrounding this question at the start of the 20th century, are best represented perhaps, by the public debate held in 1920 between Heber Curtis and Harlow Shapley. Many other authors have described this debate, in much more detail that I intend to (for example Trimble 1995), and as is well known, both astronomers had it partly right. While Curtis held the opinion that the spiral nebula represented external galaxies, his measure of the size of our Milky Way was much too small (only 10 kpc across) and he still placed the Sun at its centre. Shapley believed the Milky Way was much larger, and the Sun offset (and in this was right), but because of the vast size of our Galaxy seemed unable to conceive that the Universe could be large enough to contain millions of other similar galaxies. 

\begin{figure}
\includegraphics[height=2.5in,angle=90]{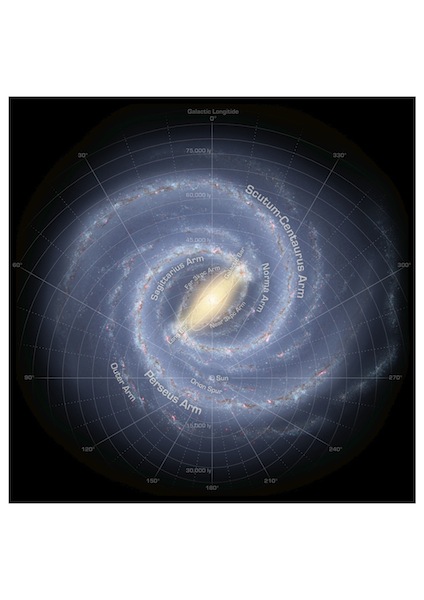}
  \caption{Modern view of the structure of our Galaxy. Image Credit: NASA/JPL-Caltech \label{fig:mw}}
\end{figure}

The modern view of the structure of our Galaxy is presented  in Figure \ref{fig:mw}, as an artists impression based on counting stars in data from the Spitzer Space Telescope observations (Benjamin 2008). Our Galaxy is a classic example of what we call a spiral galaxy, with most of its stars in a large, flat and very thin disc like structure which shows spiral arms, and a rounder region in the centre we call the ``bulge". Our Galaxy also shows evidence for an elongated ``bar" of stars across its central parts, a structure which is seen in many, but not all spiral galaxies. The Milky Way demonstrates just one of the two main kinds of big galaxies that are found in our Universe. Particularly in high density regions of the Universe, for example in the core of the nearby Virgo cluster of galaxies, many galaxies are large, smooth and spheroidal (or elliptical on the sky) in shape, and we call these types ``elliptical" galaxies. 

\section{Classification of Galaxies}

A common scientific approach to understanding a large collection of objects better is to classify it, then make categories. Much of the early effort in the field of extragalactic astronomy went into classifying the types of galaxies seen in the sky. 

Probably the most famous of the early galaxy classifiers was Edwin Hubble (1889-1953). He was not the only person at the time to develop a classification scheme, but his scheme has been the most long lasting. It has been suggested that this was because he made such broad categories, that most galaxies can fit into one of them (Buta 2011). In a defence of the scheme against criticisms made by J. H. Reynolds (Reynolds 1927), Hubble claimed to have looked at more than a thousand images of galaxies, with only a small fraction not fitting the scheme, and uncertainty in placement in ``less than ten per cent" (Hubble 1927). 

\begin{figure}
\includegraphics[height=4in,angle=0]{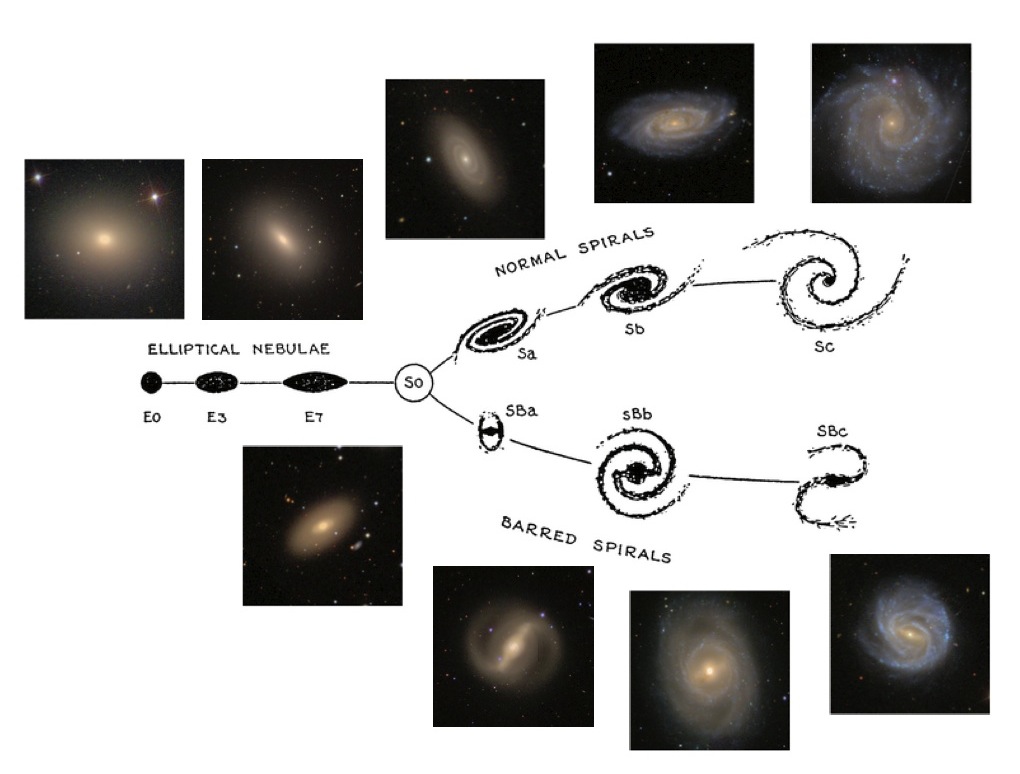}
  \caption{Hubble's original tuning fork as he presented in in Hubble (1936), with $gri$ images of galaxies from the SDSS (similar to those used in the Galaxy Zoo project) added to further illustrate each galaxy type.}\label{fig:fork}
\end{figure}

 In this same paper, Hubble described the scheme (which he first published in diagram form in {\it The Realm of the Nebula} in 1936, as shown in the centre of Figure \ref{fig:fork}). Hubble writes:  ``The classification under discussion arranges the extra-galactic nebulae in a sequence of expanding forms. There are two sections in the sequence, comprising the elliptical nebulae and the spirals respectively, which merge into one another." Ellipticals were ordered in how elliptical they appeared (from E0 being roundest, to E7 most elliptical), and spirals were ordered with reference to ``conspicuous structural features", notably (1) the size of the central bulge, (2) the extent to which the arms wound, and (3) how distinct the spiral arms were. Hubble additionally notes that the ``barred spirals" made up a distinct type of spiral ordered in the same fashion, but comprising ``a small fraction of the numbers of normal spirals" (at least in the photographic plates he used). In between the spirals and ellipticals was the (then hypothetical) S0, or lenticular type, which shows a disc, but no spiral arms. The small number of galaxies Hubble found not fitting this scheme were called ``irregular". 
 
With only minor modifications and extensions (see e.g. Buta 2011 for an extensive discussion) the Hubble classification scheme as presented is still widely used in modern extragalactic astronomy. 
We don't just keep using it because itÕs simple and it works well, but also because we have discovered that the actual physical properties of galaxies tend to vary in predictable ways along the sequence (e.g. as shown in Roberts \& Haynes 1994). It's a common misconception that Hubble labelled galaxies at the left of the diagram ``early" and those on the right ``late" in an attempt to imply an evolutionary sequence. In fact Hubble himself writes that no such conclusion should be drawn (saying that ``temporal connotations are made at one's peril" and that he set up the classification ``without prejudice to theories of [galaxy] evolution", Hubble 1927), but rather he used the labels by analogy to what was then a commonly used classification scheme for stars. These terminology are still commonly used by astronomers today to discuss relative positions of galaxies on the diagram, with ``early-type" galaxies used to mean elliptical and lenticular types, and ``late-types" for spiral galaxies. Even among spirals, early and late will be used interchangeably with the Sa, Sb, Sc notation to describe relative location along the spiral sequence.

\section{Physical Properties of Galaxies}

A galaxy is a massive collection of stars, gas and interstellar dust. To first order, the optical colour of a galaxy tells about the types of stars which live in the galaxy (ignoring the complications of dust which can redden some types of galaxies quite significantly e.g. Masters et al. 2003,  2010a).When stars form, they form in a variety of masses. The most massive stars are extremely bright, and hot and so they look blue/white (in the same way very hot metal glows blue/white). They also live for very short times (astronomically), by which we mean just a few 100 million years, and when they are visible in a galaxy the galaxy will look blue because they totally outshine everything else. 

If star formation ceases, after a short time time all the massive hot blue stars will die, and all that is left in the galaxy will be the less massive, cooler, red stars which can have lifetimes of tens of billions of years. So a galaxy which hasn't made new stars for while (more than a few 100 million years) will look red, as only red stars still shine in that galaxy. 

Back to the Hubble diagram --  the ellipticals on average are more massive (i.e. have more stars), redder, and tend to be found in clusters, while the spirals tend to be less massive, bluer and not in clusters (and increasingly so as you move along the Hubble Sequence away from the ellipticals). This tells us that the spiral galaxies are still forming stars, while the ellipticals have stopped, and more generally the Hubble sequence reveals a sequence of increasing star formation (and potential for future star formation) from left to right (Roberts \& Haynes 1994), as well as decreasing (average) total galaxy stellar mass. 

 Hubble claimed almost all galaxies he looked at fitted his sequence, however we have observations of a lot more galaxies now. Thanks to improvements in detector technology and computers able to automatically process the data, very large surveys of the sky revealing millions of galaxies have become possible, including many much fainter than those available to Hubble, and revealing extra categories of dwarf, irregular ad low surface brightness galaxies. 

\section{How Many Galaxies are There?}

In his 1927 defence of the classification scheme, Hubble claimed to have examined ``upward of a thousand galaxies" in its construction. However for true physical understanding of a galaxy, more than just an image is required -- we also need an estimate of its distance in order to reveal its size and mass. Hubble is most famous for a discovery which revolutionised extragalactic astronomy in its ability to provide a relatively quick and easy way to make these distance estimates. In 1929, Edwin Hubble used a sample of just 24 nearby galaxies to make this discovery. He used the Doppler shift of spectral lines to measure recessional velocities and used distances to these galaxies estimated based on the brightness of certain types of stars in them. By plotting these results against each other he demonstrated that the further away a galaxy is the faster it appears to be moving away from us (a finding now called ``Hubble's Law", Hubble 1929). He got the proportionality constant drastically wrong (finding what we now call Hubble's constant to be 465$\pm$50 km/s/Mpc, while around 70 km/s/Mpc is the current accepted value, e.g. as we measured in Masters et al. 2006),  but the main finding persisted, and in proceeding decades has generated an industry of using recessional velocities of galaxies (or ``redshifts" since they are revealed by a reddening of known spectral lines via the Doppler shift) to map the universe. Even to this day it remains challenging to estimate redshift independent distances to galaxies, but the spectral measurements to indicate redshifts have become relatively routine. In the first 57 years after Hubble's original findings the number of galaxies mapped had increased by a factor of 50. In 1986 the state of the art in this field was the first slice of the CfA Redshift survey (de Lapparent, Geller \& Huchra 1986). Using a sample of 1100 galaxies with redshifts this survey demonstrated that galaxies in our universe are not uniformly distributed, but rather clump in ``large scale structure".  Only 22 years after this survey, the state-of-the art was a sample 1000 times larger than this -- the final release of the SDSS Main Galaxy Sample (Strauss et al. 2002; Abazajian et al. 2009) consisting of almost 1 million galaxy redshifts. Ongoing surveys (e.g. the Baryon Oscillation Spectroscopic Survey of SDSS-III, Dawson et al. 2012) are working on increasing this even more.

\section{The ``Zoo of Galaxies" in the Era of Large Surveys}
\begin{figure}
\includegraphics[height=3in,angle=0]{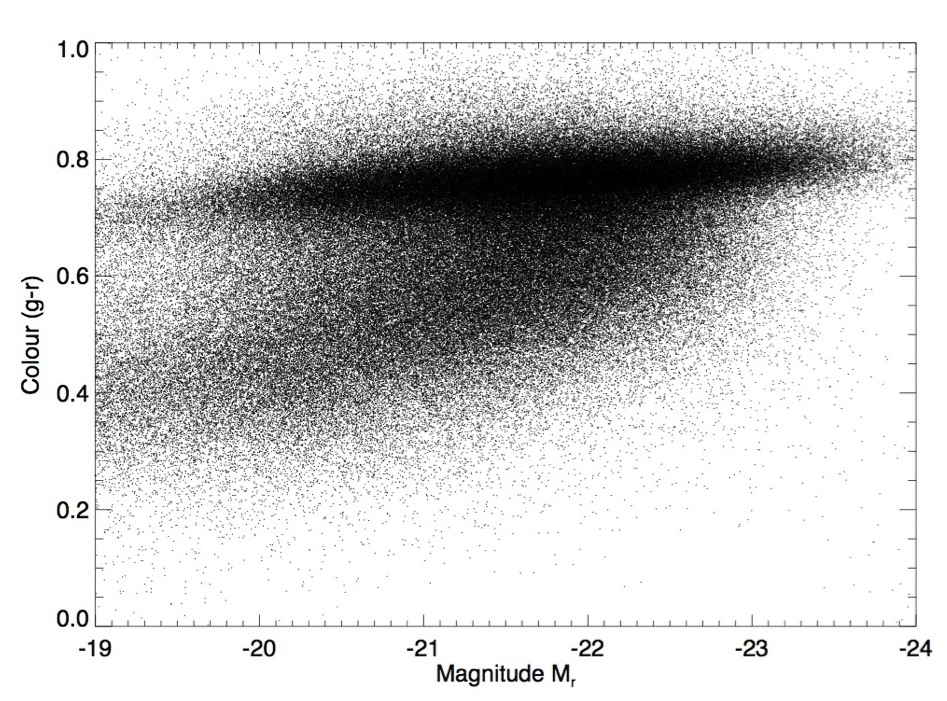}
  \caption{A colour magnitude diagram of a random selection of the SDSS Main Galaxy Sample (Strauss et al. 2002). This plots the $(g-r)$ colour versus the $r$-band absolute magnitude of almost 300,000 randomly selected galaxies in the Data Release 6 version (Adelman-McCarthy et al. 2008). }\label{fig:cm}
\end{figure}

Modern large surveys of hundreds of thousands of galaxies with detailed images and spectra (like the SDSS Main Galaxy Sample) have inspired a new field in extragalactic astronomy -- the use of statistical analyses to reveal demographics of the population of galaxies. Perhaps the most famous example of this in the field of extragalactic astronomy is the use of the colour-magnitude diagram as a basic observation of galaxies (e.g. Strateva et al. 2001, Bell et al. 2003, Kauffmann et al. 2003, Baldry et al. 2004, Balogh et al. 2004). The galaxy population show a striking bimodality in this diagram, with galaxies mostly found in two regions, which have become known as the ``red sequence" and the ``blue cloud", with a sparsely populated ``green valley" in between (e.g. a fraction of the galaxies of the SDSS Main Galaxy Sample are shown on this diagram in Fig \ref{fig:cm}). This diagram demonstrates a general trend among galaxies that bigger (or intrinsically brighter) galaxies tend to be optically redder (ie. having an older stellar population). This trend is most striking in the red sequence, but also apparent in the blue cloud. Hubble Space Telescope surveys have demonstrated that this colour magnitude diagram is in place since at least $z\sim1$ (Bell et al. 2004). They also show that at earlier times in Universe more galaxies were found in the blue cloud -- demonstrating that on average galaxies move from blue to red as cosmic time progresses. Much of extragalactic astronomy in recent decades has focused on developing an understanding of the mechanisms which shape the locations galaxies are found in diagram and the physical processes which move them around on it.

The bimodality in colour mirrors the two galaxy types discussed above, with the blue cloud hosting mostly spiral galaxies, and the red sequence mostly elliptical and lenticular galaxies. What had been missing in making this statement though was morphology (or type) for more than a few thousand objects (e.g. Fukugita et al. 2007 classified $\sim$2000 of the SDSS galaxies). It seemed impossible to do that for the sample sizes being generated by astronomers in the early part of the 21st century (although the MOSES project tried -- visually inspecting 50,000 galaxies to search for blue ellipticals, Schawinski et al. 2007). Without reliable information on galaxy morphology however, it's not possible to have the full picture of galaxy evolution. The colour of a galaxy is driven by the stellar (and gas and dust) content of the galaxy, while the shape or morphology of a galaxy reflects its dynamical history which could be very different (and have a different timescale). Therefore, one of the central motivations for the original Galaxy Zoo project was to construct a large sample of early and late type galaxy classifications from SDSS that were independent of colour.

\section{Galaxy Zoo}
Galaxy Zoo in its original form was launched on July 11, 2007 and introduced in a BBC online article that same day\footnote{{\it Scientists seek galaxy hunt help}, by Christine McGourty \\ ({\tt http://news.bbc.co.uk/1/hi/sci/tech/6289474.stm})}.  The original site simply asked volunteers to classify galaxies as either spiral or elliptical -- the most basic morphological split among galaxies. Something in Galaxy Zoo resonated extraordinarily with the general public. The original projection estimated that if a few thousand members of the public got involved the 1 million galaxies might be classified in a couple of years. However, within twelve hours of the launch, the Galaxy Zoo site was receiving 20,000 classification per hour. After forty hours, the classification rate had increased to 60,000 per hour. After ten days, the public had submitted $\sim8$ million classifications.  By April 2008, when the Galaxy Zoo team submitted their first paper (Lintott et al 2008), over 100,000 volunteers had classified each of the $\sim900,000$ SDSS galaxy images an average of $38$ times. 

The popularity of Galaxy Zoo and the number of classifications received enabled science which simply would not have been possible without its contribution. Not only does Galaxy Zoo give a classification for each galaxy, but also by collecting $\sim 40$ independent classifications of each galaxy Galaxy Zoo produces an estimate of how likely that classification is to be be true (a classification ``likelihood"). All of the data from this first phase of Galaxy Zoo were published in Lintott et al. (2011) and are available to download via the SDSS servers\footnote{\tt http://skyserver.sdss3.org/CasJobs/}. For an in-depth history and analysis of the Galaxy Zoo project see Fortson et al. (2012).

\subsection{Red Spirals}

 One of the biggest contributions of Galaxy Zoo has been in finding large samples of rare classes of extragalactic object in the SDSS sample. An example of this are the ``red spirals" --  relatively rare spiral galaxies which are found on the red sequence (first discussed in Bamford et al. 2009, Skibba et al. 2009). Due to the reddening effects of dust, a significant number of edge-on spirals are found in the red sequence, and spiral galaxies with large central bulges can also be intrinsically very red due to old stars in the bulge (Masters et al. 2010a), however Galaxy Zoo revealed a signifiant fraction of even late-type spirals found on the red sequence (Masters et al. 2010b). These are galaxies which have small bulges, but intrinsically red discs yet still show clear spiral arms. The provide a direct probe of evolution affecting star formation but not morphology, revealing that processes exist which can turn spiral galaxies red without disturbing their morphology. Even though these are relatively rare objects in the galaxy population, studies have suggested they form a signifiant part of the route for most of the evolution from the blue cloud to the red sequence both with increasing density in the local universe (Bamford et al. 2009) and with redshift (Bundy et al. 2010). In Masters et al. (2010b) we revealed that while these objects are most common in intermediate density regions (as also shown in Bamford et al. 2009 and Skibba et al. 2009) they are found even at very low densities. We demonstrated that compared to blue cloud spirals, red spirals become more common as spiral galaxies become more massive, that they are not significantly more dusty (as revealed by Balmer decrements) but are significantly more likely to host LINER (Low Ionization Nuclear Emission Region) like emission and obvious bars. The Galaxy Zoo red spirals are not completely passive, but at fixed stellar mass show significantly less star formation and an older stellar population than their blue cousins. 

 The red spirals have provided part of the evidence prompting a move from the view of major mergers as the main route of galaxy evolution (e.g. as presented in Steinmetz \& Navarro 2002), to more gentle and slower (secular) processes playing a significant role for most galaxies.  In order to make a spiral galaxy red (by stopping star formation), something has happened to these galaxies to exhaust, or remove their supply of atomic hydrogen gas (the raw material for star formation). It's been discussed for many years that in our galaxy the amount of gas in the disc will be used up by the current rate of star formation in much less than a Hubble time (Larson, Tinsley \& Caldwell 1980), and evidence exists that this gas is being replenished by the infall and cooling of hot gas from the  halo (for a review see Putman et al. 2012). If this gas supply is shut off (either by being removed, or heated by processes like strangulation or harassment, Larson, Tinsley \& Caldwell 1980, Balogh, Navarro \& Morris 2000, Bekki et al. 2002) star formation will cease and the disc redden on a timescale of $\sim1$ Gyr. In fact recent work modelling the star formation histories of the Galaxy Zoo red spirals compared to blue spirals and red ellipticals supports this picture, showing that the average star formation history of red spirals only differs from blue spirals in the last 1 Gyr (Tojeiro et al. 2012).
 
 One of the most striking observations about the red spirals was that such a large fraction of them were very strongly barred (75\% Masters et al. 2010b). In much of the study of galaxy evolution, the division between barred and unbarred spirals has been ignored. It has often been argued that these types of galaxies are intrinsically the same, just caught with or without a bar (e.g. as recently discussed by van den Bergh 2011), even though most theoretical considerations now suggest bars must be very long lived. Around 30-60\% of massive spirals host bars, with the exact value depending on how strict you are with the definition of what a bar is, which can vary from a linear feature stretching across most of the galaxy (as in the classic example NGC 1300), to mild oval distortions in the central regions.  

\subsection{Galaxy Zoo 2 and Barred Spirals} 
 
\begin{figure}
\includegraphics[height=3in,angle=0]{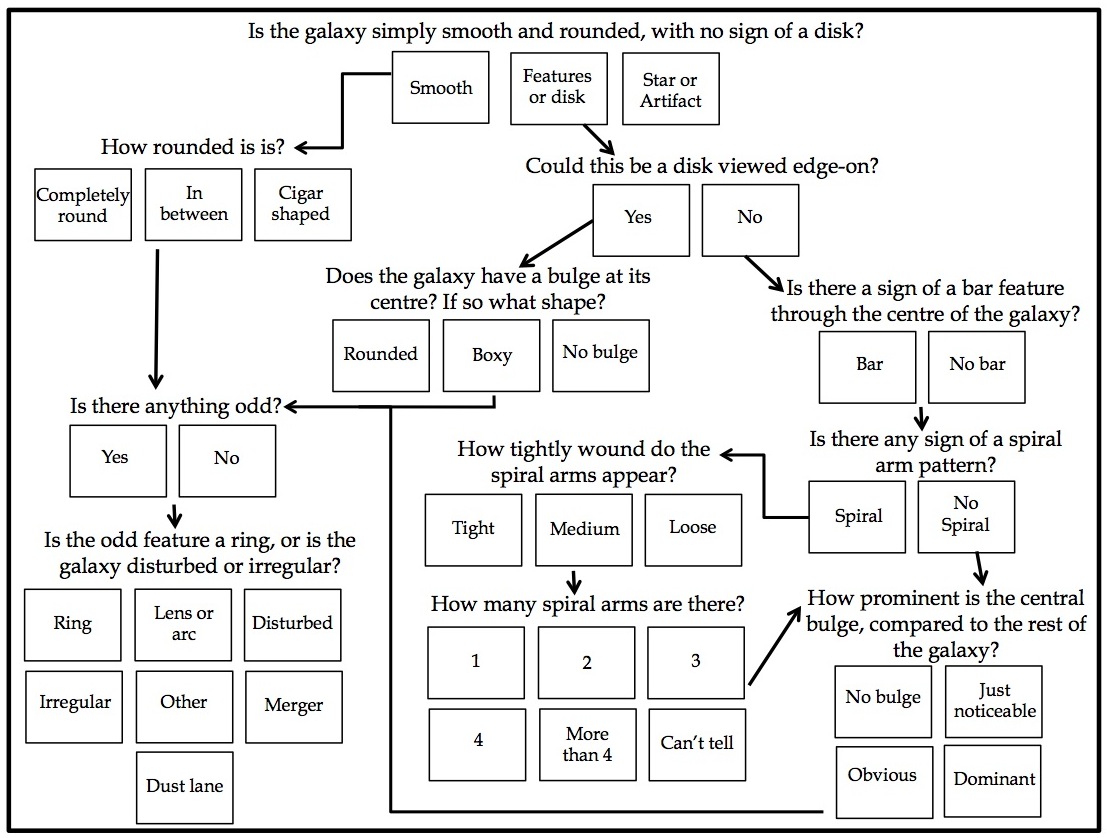}
  \caption{Full classification tree from Galaxy Zoo 2. First published in Masters et al. (2011).}\label{fig:gz2}
\end{figure}

 At the time we noticed the large bar fraction in red spirals, data from the second phase of Galaxy Zoo was starting to become available. Galaxies can show all sorts of interesting structures beyond the simple split between spiral and elliptical, and these features reveal more clues to the formation histories of the galaxies. Galaxy Zoo was so popular, and the results so reliable (agreeing with experts just as well as experts are able to agree with each other, as discussed in Lintott et al. 2008) that a new version was developed, asking for significantly more detail for the brightest quarter of the original sample. If a galaxy was identified as being a disk or showing features, this version asked questions about the number of spiral arms, size of the bulge, and most importantly for us, the presence of a bar. The full classification scheme for Galaxy Zoo 2 (GZ2) is shown in Figure \ref{fig:gz2}, reproduced from Masters et al. (2011). This version of the site launched in February 2009 and ran for fourteen months, collecting in that time 60 million individual classifications of the images. 

The first science result from GZ2 were a study of the bar fraction of disc galaxies as a function of colour. This work (Masters et al. 2011) demonstrated that the types of bars revealed by GZ2\footnote{which have been shown to be very similar to the classical ``strong" bars identified in early visual catalogues - see Masters et al. (2012) for a discussion.} were significantly more common in redder, more massive disc galaxies -- with the extreme of this being the red spirals, among with 75\% showed bars. This work also revealed a split in bar fraction between disc galaxies with large and small bulges (as revealed by fits to the light profile from SDSS) such that bulge dominated spirals showed more bars. These types of trends of bar fraction with galaxy colour have also been seen in the local universe by Nair \& Abraham (2010) and Giordarno et al. (2010) and at higher redshift by Sheth et al. (2008). The GZ2 results also hinted at an upturn of the bar fraction in the bluest and lowest mass disc galaxies as observed more clearly by Barazza et al. (2009) and Aguerri et al. (2009) in disc galaxy samples dominated by these kinds of galaxies.  

 The striking thing about our theoretical understanding of bars in disc galaxies, is that forming a bar in a disc galaxy is extremely easy. The question is not why some galaxies make them, but why some do not (or have not yet). 
 
  While a bar is forming it enables the exchange of angular momentum in a disc galaxy -- basically moving material around. This has the effect of growing central concentrations (Kormendy \& Kennicutt 2004), sparking central star formation (e.g. as observed by Sheth et al. 2005, Ellison et al. 2011), and possibly helping to feed central active galactic nuclei (although this is more controversial e.g. Ho et al. 1997, or Oh et al. 2012, Cardamone et al. in prep.). Theoretical considerations indicate that the gas content of a galaxy is important in this process (e.g. Athanassoula 2003; Combes 2008). By virtue of being able to dissipate energy, gas is an important sink of angular momentum, and any galaxy with a significant gas fraction in simulations struggles to develop a large bar (e.g. Villa-Vargas et al. 2010), while in a simulation of a gas poor disc, bars can be very stable and long lived. 
  
  The effective forces produced by the bar instability act to drive gas inwards from co-rotation (the point at which stars in the disk rotate with the same speed as the pattern speed of the bar) to the central regions. This gas looses its angular momentum which is transfered to the stars in the bar. Interestingly, the forces outside the ends of the bar may also act to inhibit inflow of gas from the outer regions of the disc, so that gas inflow of external gas onto a disc galaxy is inhibited in the presence of a strong bar (Combes 2008). It is possible that this process could play a role in the global reddening of disc galaxies, and while the concentration of gas in the centres of galaxies caused by the bar will spark central star formation, it evacuates gas from the rest of the disc (preventing star formation there) and potentially helps to use up the total gas content of the galaxy quicker. 

\subsection{Bars and Atomic Gas Content}

  The discovery that the type of strong, or obvious bars easily identified in GZ2 classifications are more common in more massive, redder disc galaxies, already implies that bars are more often found in gas poor disc galaxies, since optical colour is a good proxy for current star formation rate, which correlates well with gas content. However we wanted to investigate more directly the role of gas content on the likelihood of a galaxy hosting a bar. 
  
 The Arecibo Legacy Fast ALFA (Arecibo L-band Feed Array) survey (ALFALFA; Giovanelli et al. 2005) provide the ideal data to do this, matched well to the available GZ2 bar sample. ALFALFA has mapped all of the high Galactic latitude sky visible from the Arecibo Radio Telescope (which is located in Puerto Rico) in the 21cm line emitted by neutral hydrogen atoms (HI). This is giving an amazing census of the atomic hydrogen gas content of all galaxies in this part of the sky, which overlaps much of the SDSS Legacy Imaging area in which Galaxy Zoo galaxies are selected. In Masters et al. (2012) we constructed a sample of 2090 disc galaxies with bar identifications from GZ2 and neutral hydrogen detections from the 40\% of ALFALFA available at the time (Haynes et al. 2011). Our main result was that we confirmed observationally that bars are more likely to be found in galaxies with less gas. In fact we showed that this is also true in galaxies which have less gas than is typical for their stellar mass and optical colour, and perhaps more interestingly that disc galaxies with both gas and bars are optically redder than similar disc galaxies with gas and no bar -- perhaps the first direct evidence that the bar could be helping to globally redden its host galaxy. 
 
 This is a work in progress, with many more interesting results coming out on the GZ2 bars. For example in Hoyle et al. (2011) we demonstrated that redder disc galaxies not only host more bars, but that these bars are also longer (and therefore stronger). In Skibba et al. (2012) we provided the first clear evidence for an environmental effect on bar formation, finding more bars in galaxies clustered on certain scales even when corrected for galaxy mass and colour. Casteels et al. (2012) shows that in close pairs bar formation is suppressed, and Cardamone et al. (in prep.) will show that once corrected for galaxy mass, we observe no correlation between active galactic nuclei and bars identified in GZ2. All of this is providing vital clues to reveal the role of bars on the global evolution of disc galaxies. 
 
\section{Whats Next for the Zoo?}\label{sec:concl}

The popularity of Galaxy Zoo was both immediate and long lasting, and raised questions very early on with the science team about why so many people would choose to spend so much time classifying galaxies online. A survey was launched to study the motivations of the citizen scientists participating in Galaxy Zoo, which was published by Raddick et al. (2010). For a scientist involved in research using the Galaxy Zoo classifications, the most striking things about this survey was that so many people identified a desire to help with scientific research as their main motivation. This means that the scientific results coming out of Galaxy Zoo are essential to the ongoing success of the project in attracting and retaining volunteer classifiers. Fortunately Galaxy Zoo was designed with specific and immediate science questions to answer. The first publications from Galaxy Zoo came out within a year of launch (Lintott et al. 2008, Land et al. 2008), and the total number of peer reviewed papers based on Galaxy Zoo data is now more than 30, with a growing number of results from scientists not involved in the Galaxy Zoo project, and using the publicly released GZ1 data. The classifications from GZ2 are also in the process of being prepared for public release. A selection of papers from the Galaxy Zoo team are listed in Table 1. 

\begin{table}
\caption{\label{gz1papers} A selection of peer reviewed papers based on classifications collected from Galaxy Zoo (in order of publication).}
{\scriptsize \begin{tabular}{ll}
Author \& Year & Title - Galaxy Zoo:  \\
\hline
Land et al. 2008  & The large-scale spin statistics of spiral galaxies in the Sloan Digital Sky Survey \\
Lintott et al.  2008  & Morphologies derived from visual inspection of   galaxies from the SDSS \\
Slosar et al.  2009  & Chiral correlation function of galaxy spins \\
Bamford et al. 2009 & The dependence of morphology and colour on environment \\
Schawinski et al. 2009a & A sample of blue early-type galaxies at low redshift \\
Lintott et al. 2009  & `Hanny's Voorwerp', a quasar light echo? \\
Skibba et al. 2009  & Disentangling the environmental dependence of morphology and colour \\
 Cardamone et al.  2009  & Green Peas: discovery of a class of compact extremely star-forming galaxies \\
 Darg et al.  2010a  & The fraction of merging galaxies in the SDSS and their morphologies \\
 Darg et al. 2010b  & The properties of merging galaxies in the nearby  Universe - local environments, \\& colours, masses,  star formation rates and AGN activity \\
 Schawinski et al. 2010a  & The Sudden Death of the Nearest Quasar \\
 Schawinski et al. 2010b  & The Fundamentally Different Co-Evolution of  Supermassive Black Holes and \\& Their Early- and  Late-Type Host Galaxies\\
 Masters et al.  2010a  & Dust in spiral galaxies\\
 Jimenez et al. 2010  & A correlation between the coherence of galaxy spin  chirality and star formation efficiency \\
Masters et al. 2010b  & Passive red spirals \\
 Banerji et al,  2010 & Reproducing galaxy morphologies via machine learning \\
 Lintott et al 2011  & Data Release of Morphological Classifications for  nearly 900,000 galaxies \\
 Masters et al. 2011 & Bars in disc galaxies \\
 Smith et al. 2011 & Galaxy Zoo Supernovae \\
 Hoyle et al. 2011 & Bar lengths in local disc galaxies \\
 Darg et al. 2011 &  Multi-Mergers and the Millennium Simulation \\
 Keel et al. 2012 & The Galaxy Zoo survey for giant AGN-ionized clouds: past and present black hole \\&  accretion events\\
 Wong et al. 2012  & Building the low-mass end of the red sequence with  local post-starburst galaxies \\
 Kaviraj et al. 2012 & Dust and molecular gas in early-type galaxies with prominent dust lanes\\
 Shabala et al. 2012 & Dust lane early-type galaxies are tracers of recent, gas-rich minor mergers\\
 Skibba et al. 2012 & The environmental dependence of bars and bulges in disc galaxies \\
 Masters et al. 2012 & Atomic gas and the regulation of star formation in barred disc galaxies\\
 Hoyle et al. 2012 & The fraction of early-type galaxies in low-redshift groups and clusters of galaxies \\
 Teng et al. 2012 & Chandra Observations of Galaxy Zoo Mergers: Frequency of Binary Active Nuclei in \\& Massive Mergers\\
 Simmons et al. 2012 & Bulgeless Galaxies With Growing Black Holes\\
 Casteels et al. 2012 & Quantifying Morphological Indicators of Galaxy Interaction\\
\end{tabular}}
\end{table}

Galaxy Zoo was a pioneer in what has becoming a new methodology of involving ``citizen scientists" in research. The Zooniverse\footnote{\tt www.zooniverse.org} was launched in December 2009 on the back of the success of Galaxy Zoo, and provides a framework for collecting a variety of similar projects. This umbrella now includes several other astronomically themed projects (e.g. Milky Way Project, Moon Zoo, Supernova Zoo, Solar Stormwatch) but also projects in other areas of science, such as Old Weather (extracting climate data from handwritten ship logbooks), Whale.FM (comparing whale song), Cell Slider (identifying cancer cells in images) and even non science projects (e.g. Ancient Lives, which attempts to piece together scraps of papyrus). All of these and more can be accessed via the main Zooniverse website. 

The Zooniverse team are also working on tools to improve the educational use of these sites. They provide a way to involve school children in real hands on science, but working out how to include this in lesson plans has proved difficult. Recently launched was Galaxy Zoo: Navigator\footnote{\tt www.galaxyzoo.org/\#/navigator} which allows a group to collect their classifications together, and use online tools to explore how these depend on other galaxy properties (from SDSS data). This increases the learning potential from Galaxy Zoo. The Zooniverse also hosts a website called ``Zooteach"\footnote{\tt www.zooteach.org/} for teachers and educators to share lesson plans and ideas for the use of Zooniverse projects in the classroom. 

Galaxy Zoo itself remains one of the most successful and popular of the Zooniverse sites. Following GZ2, Galaxy Zoo: Hubble ran (April 2010-September 2012) collecting classifications on galaxies observed by the Hubble Space Telescope (e.g. as part of COSMOS, GOODS, EDS and other large surveys). Shortly after this Invited Discourse, Galaxy Zoo relaunched in its fourth version, now including images of galaxies from the HST CANDELS survey as well as new images of galaxies from the SDSS-III imaging area. 

\begin{figure}
\includegraphics[width=3in]{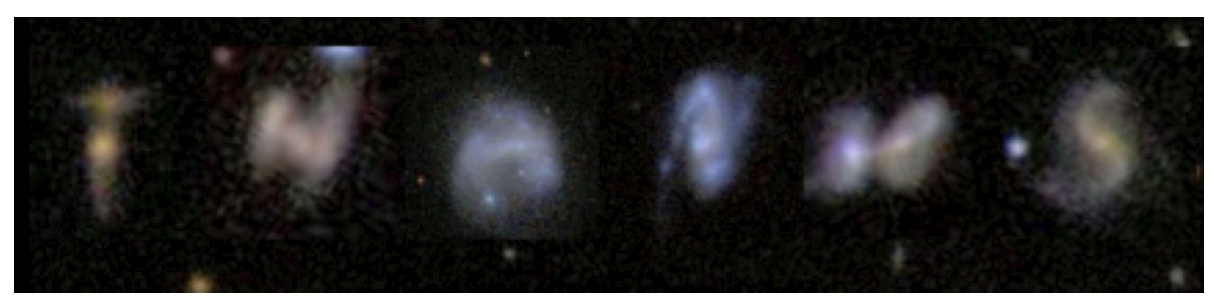}
  \caption{``Thanks" written in the galaxy alphabet found by Galaxy Zoo volunteers. Write your own words at  {\tt writing.galaxyzoo.org}}\label{fig:thanks}
\end{figure}

 I'll finish by just saying thank you to all of the more than 200,000 volunteers who have helped make galaxy classifications via the Galaxy Zoo website by using the special ``galaxy font" they helped make (Fig \ref{fig:thanks})\footnote{Available at {\tt writing.galaxyzoo.org}}.  To anyone who has not yet tried out Galaxy Zoo, I'd like to encourage you to come and play in our ``X\={i}ng X\`{i} Z\v{o}ng D\`{o}ng Yu\'{a}n" (``Galaxy Zoo" in Chinese). 

\begin{acknowledgments}
I'd like to thank the organisers of the 28th GA of the IAU for inviting me to give this talk. 

This publication has been made possible by the participation of more than 200,000 volunteers in the Galaxy Zoo project. Their contributions are individually acknowledged at http://www.galaxyzoo.org/volunteers. 

I acknowledge funding from the Peter and Patricia Gruber Foundation as the 2008 Peter and Patricia Gruber Foundation IAU Fellow,  and from a 2010 Leverhulme Trust Early Career Fellowship, as well as support from the Royal Astronomical Society to attend the 28th GA of the IAU.

Thanks to Dongai for helping with my Chinese introductory remarks to the Invited Discourse. 
\end{acknowledgments}


\begin{thebibliography}{}
\bibitem[Abazajian et al.(2009)]{2009ApJS..182..543A} Abazajian, K.~N., Adelman-McCarthy, J.~K., Ag{\"u}eros, M.~A., et al.\ 2009, ApJS, 182, 543 
\bibitem[Adelman-McCarthy et al.(2008)]{2008ApJS..175..297A} Adelman-McCarthy, J.~K., Ag{\"u}eros, M.~A., Allam, S.~S., et al.\ 2008,ApJS, 175, 297 
\bibitem[Aguerri et al. (2009)]{Ag09} Aguerri, J.~A.~L., M{\'e}ndez-Abreu, J., \& Corsini, E.~M.\ 2009, A\&A, 495, 491 
\bibitem[Aihara et al.(2011)]{2011ApJS..193...29A} Aihara, H., Allende Prieto, C., An, D., et al.\ 2011, ApJS, 193, 29 
\bibitem[Athanassoula(2003)]{2003MNRAS.341.1179A} Athanassoula, E.\ 2003, MNRAS, 341, 1179 
\bibitem[Baldry et al.(2004)]{2004ApJ...600..681B} Baldry, I.~K., Glazebrook, K., Brinkmann, J., Ivezi{\'c}, {\v Z}., Lupton, R.~H., Nichol, R.~C., \& Szalay, A.~S.\ 2004, ApJj, 600, 681 
\bibitem[Bell et al.(2003)]{2003ApJS..149..289B} Bell, E.~F., McIntosh, D.~H., Katz, N., \& Weinberg, M.~D.\ 2003, ApJS, 149, 289 
\bibitem[Bell et al.(2004)]{2004ApJ...608..752B} Bell, E.~F., Wolf, C., Meisenheimer, K., et al.\ 2004, ApJ, 608, 752 
\bibitem[Balogh et al.(2000)]{2000ApJ...540..113B} Balogh, M.~L., Navarro, J.~F., \& Morris, S.~L.\ 2000, ApJ, 540, 113 
\bibitem[Balogh et al.(2004)]{2004ApJ...615L.101B} Balogh, M.~L., Baldry, I.~K., Nichol, R., et al.\ 2004, ApJL, 615, L101 
\bibitem[Bamford et al.(2009)]{B09} Bamford, S.~P., et al.\ 2009, MNRAS, 393, 1324
\bibitem[Banerji et al.(2010)]{2010MNRAS.406..342B} Banerji, M., Lahav, O., Lintott, C.~J., et al.\ 2010, MNRAS, 406, 342 
\bibitem[Barazza et al. (2008)]{Bar08} Barazza, F.~D., Jogee, S., \& Marinova, I.\ 2008, ApJ, 675, 1194 
\bibitem[Bekki et al.(2002)]{Bekki02} Bekki, K., Couch, W.~J., \& Shioya, Y.\ 2002, ApJ, 577, 651 
\bibitem[Benjamin(2008)]{2008ASPC..387..375B} Benjamin, R.~A.\ 2008, Massive Star Formation: Observations Confront Theory, 387, 375 
\bibitem[Bundy et al.(2010)]{B10} Bundy, K., et al.\ 2010, ApJ, 719, 1969 
\bibitem[Buta(2011)]{2011arXiv1102.0550B} Buta, R.~J.\ 2011, in {\it Planets, Stars, and Stellar Systems}, Vol. 6, Series Editor T. D. Oswalt, Volume editor W. C. Keel, Springer (arXiv:1102.0550)
\bibitem[Cardamone et al.(2009)]{2009MNRAS.399.1191C} Cardamone, C., Schawinski, K., Sarzi, M., et al.\ 2009, MNRAS, 399, 1191 
\bibitem[Casteels et al.(2012)]{2012arXiv1206.5020C} Casteels, K.~R.~V., Bamford, S.~P., Skibba, R.~A., et al.\ 2012, MNRAS (submitted: arXiv:1206.5020)
\bibitem[Combes (2008)]{2008IAUS..245..151C} Combes, F.\ 2008, in {\it Formation and Evolution of Galaxy Bulges}, Proceedings of IAU Symposium, 245, Eds M. Bureau, E. Athanassoula and B. Barbuy
\bibitem[Darg et al.(2010)]{2010MNRAS.401.1043D} Darg, D.~W., Kaviraj, S., Lintott, C.~J., et al.\ 2010, MNRAS, 401, 1043 
\bibitem[Darg et al.(2010)]{2010MNRAS.401.1552D} Darg, D.~W., Kaviraj, S., Lintott, C.~J., et al.\ 2010, MNRAS, 401, 1552 
\bibitem[Darg et al.(2011)]{2011MNRAS.416.1745D} Darg, D.~W., Kaviraj, S., Lintott, C.~J., et al.\ 2011, MNRAS, 416, 1745 
\bibitem[Dawson et al.(2012)]{2012arXiv1208.0022D} Dawson, K.~S., Schlegel, D.~J., Ahn, C.~P., et al.\ 2012, AJ (submitted; arXiv:1208.0022)
\bibitem[de Lapparent et al.(1986)]{1986ApJ...302L...1D} de Lapparent, V., Geller, M.~J., \& Huchra, J.~P.\ 1986, ApJL, 302, L1 
\bibitem[Ellison et al.(2011)]{2011MNRAS.416.2182E} Ellison, S.~L., Nair, P., Patton, D.~R., et al.\ 2011, MNRAS, 416, 2182  
\bibitem[Fortson et al.(2012)]{2012amld.book..213F} Fortson, L., Masters, K., Nichol, R., et al.\ 2012, Advances in Machine Learning and Data Mining for Astronomy, CRC Press, Taylor \& Francis Group, Eds.: Michael J.~Way, Jeffrey D.~Scargle, Kamal M.~Ali, Ashok N.~Srivastava, p.~213-236, 213  (arXiv:1104.5513)
\bibitem[Giordano et al. (2012)]{G10} Giordano, L., Tran, K.V.H., Moore, B., \& Saintonge, A. 2012, ApJ (submitted; arXiv:1002.3167) 
\bibitem[Giovanelli et al.(2005)]{G05} Giovanelli, R., et al.\ 2005, AJ, 130, 2598 
\bibitem[Haynes et al.(2011)]{2011AJ....142..170H} Haynes, M.~P., Giovanelli, R., Martin, A.~M., et al.\ 2011, AJ, 142, 170 ($\alpha40$) 
\bibitem[Herschel(1785)]{H1785} Herschel, W. 1785, Philosophical Transactions of the Royal Society of London, Vol. 75., 213-266
\bibitem[Ho et al.(1997)]{1997ApJ...487..591H} Ho, L.~C., Filippenko,  A.~V., \& Sargent, W.~L.~W.\ 1997, ApJ, 487, 591 
\bibitem[Hoyle et al.(2011)]{2011MNRAS.415.3627H} Hoyle, B., Masters, K.~L., Nichol, R.~C., et al.\ 2011, MNRAS, 415, 3627 
\bibitem[Hoyle et al.(2012)]{2012MNRAS.423.3478H} Hoyle, B., Masters, K.~L., Nichol, R.~C., Jimenez, R., \& Bamford, S.~P.\ 2012, MNRAS, 423, 3478 
\bibitem[Hubble(1927)]{1927Obs....50..276H} Hubble, E.~P.\ 1927, The Observatory, 50, 276 
\bibitem[Hubble(1936)]{1936rene.book.....H} Hubble, E.~P.\ 1936, {\it Realm of the Nebulae}, by E.P.~Hubble.~ New Haven: Yale University Press, 1936.~ ISBN 9780300025002,  
\bibitem[Jimenez et al.(2010)]{2010MNRAS.404..975J} Jimenez, R., Slosar, A., Verde, L., et al.\ 2010, MNRAS, 404, 975 
\bibitem[Kauffmann et al.(2003)]{2003MNRAS.341...54K} Kauffmann, G., Heckman, T.~M., White, S.~D.~M., et al.\ 2003, MNRAS, 341, 54 
\bibitem[Kaviraj et al.(2012)]{2012MNRAS.423...49K} Kaviraj, S., Ting, Y.-S., Bureau, M., et al.\ 2012, MNRAS, 423, 49 
\bibitem[Keel et al.(2012)]{2012MNRAS.420..878K} Keel, W.~C., Chojnowski, S.~D., Bennert, V.~N., et al.\ 2012, MNRAS, 420, 878 
\bibitem[Kormendy \& Kennicutt(2004)]{KK04} Kormendy, J., \& Kennicutt, R.~C., Jr.\ 2004, ARA\&A, 42, 603 
\bibitem[Land et al.(2008)]{La08} Land, K., Slosar, A., Lintott, C.J.,  et al.\ 2008, MNRAS, 388, 1686 
\bibitem[Larson et al.(1980)]{L80} Larson, R.~B., Tinsley, B.~M., \& Caldwell, C.~N.\ 1980, ApJ, 237, 692 
\bibitem[Lintott et al.(2008)]{L08} Lintott, C.~J., et al.\ 2008, MNRAS, 389, 1179 
\bibitem[Lintott et al.(2009)]{2009MNRAS.399..129L} Lintott, C.~J., Schawinski, K., Keel, W., et al.\ 2009, MNRAS, 399, 129 
\bibitem[Lintott et al.(2011)]{L11} Lintott, C.~J., et al.\ 2011, MNRAS, 410, 166
\bibitem[Masters et al. (2003)]{M03} Masters, K. L., Giovanelli, R. \& Haynes, M. P. 2003, AJ, 126, 158. 
\bibitem[Masters et al. (2006)]{M06} Masters, K. L., Haynes, M. P., Giovanelli, R. \&  Springob, C. M. 2006, ApJ 653, 861 
\bibitem[Masters et al. (2010a)]{GZdust} Masters, K. L., et al. 2010a, MNRAS ~404, 792. 
\bibitem[Masters et al. (2010b)]{M10} Masters, K. L., et al. 2010b, MNRAS ~405, 783.  
\bibitem[Masters et al. (2011)]{M11} Masters, K. L., et al. 2011, MNRAS, 411, 2026 
\bibitem[Masters et al.(2012)]{2012MNRAS.424.2180M} Masters, K.~L., Nichol, R.~C., Haynes, M.~P., et al.\ 2012, MNRAS, 424, 2180 
\bibitem[Messier(1781)]{1781cote.rept..227M} Messier, C.\ 1781, Connoissance des Temps for 1784, p.~227-267, 227 
\bibitem[Nair \& Abraham(2010b)]{NA10} Nair, P.~B., \& Abraham, R.~G.\ 2010b, ApJ L, 714, L260 
\bibitem[Oh et al.(2012)]{2012ApJS..198....4O} Oh, S., Oh, K., \& Yi, S.~K.\ 2012, ApJs, 198, 4 
\bibitem[Putman et al.(2012)]{2012ARA&A..50..491P} Putman, M.~E., Peek, J.~E.~G., \& Joung, M.~R.\ 2012, ARA\&A, 50, 491 
\bibitem[Raddick et al.(2010)]{2010AEdRv...9a0103R} Raddick, M.~J., Bracey, G., Gay, P.~L., et al.\ 2010, Astronomy Education Review, 9, 010103 
\bibitem[Roberts \& Haynes(1994)]{1994ARA&A..32..115R} Roberts, M.~S., \& Haynes, M.~P.\ 1994, ARA\&A, 32, 115 
\bibitem[Schawinski et al.(2007)]{S07} Schawinski K., Thomas D., Sarzi M., Maraston C., Kaviraj S., Joo S.-J., Yi S.~K., Silk J.\ 2007, MNRAS, 382, 1415
\bibitem[Schawinski et al.(2009)]{2009MNRAS.396..818S} Schawinski, K., Lintott, C., Thomas, D., et al.\ 2009, MNRAS, 396, 818 
\bibitem[Schawinski et al.(2010a)]{2010ApJ...724L..30S} Schawinski, K., Evans, D.~A., Virani, S., et al.\ 2010a, ApJL, 724, L30 
\bibitem[Schawinski et al.(2010b)]{2010ApJ...711..284S} Schawinski, K., Urry, C.~M., Virani, S., et al.\ 2010b, ApJ, 711, 284 
\bibitem[Simmons et al.(2012)]{2012arXiv1207.4190S} Simmons, B.~D., Lintott, C., Schawinski, K., et al.\ 2012, MNRAS (submitted; arXiv:1207.4190)
\bibitem[Shabala et al.(2012)]{2012MNRAS.423...59S} Shabala, S.~S., Ting, Y.-S., Kaviraj, S., et al.\ 2012, MNRAS, 423, 59 
\bibitem[Sheth et al.(2005)]{S05} Sheth, K., Vogel, S.~N., Regan, M.~W., Thornley, M.~D., \& Teuben, P.~J.\ 2005, ApJ, 632, 217
\bibitem[Sheth et al.(2008)]{S07} Sheth, K., et al.\ 2008, ApJ, 675, 1141 
\bibitem[Slosar et al.(2009)]{2009MNRAS.392.1225S} Slosar, A., Land, K., Bamford, S., et al.\ 2009, MNRAS 392, 1225 
\bibitem[Skibba et al.(2009)]{Sk09} Skibba, R.~A., et al.\ 2009, MNRAS, 399, 966
\bibitem[Skibba et al.(2012)]{2012MNRAS.tmp.2854S} Skibba, R.~A., Masters, K.~L., Nichol, R.~C., et al.\ 2012, MNRAS, 423, 1485
\bibitem[Smith et al.(2011)]{2011MNRAS.412.1309S} Smith, A.~M., Lynn, S., Sullivan, M., et al.\ 2011, MNRAS, 412, 1309 
\bibitem[Steinicke(2012)]{S12} Steinecke, W. 2012, Journal and Astronomical History and Heritage, 15(1), 19-29.
\bibitem[Steinmetz \& Navarro(2002)]{2002NewA....7..155S} Steinmetz, M., \& Navarro, J.~F.\ 2002, New Astronomy, 7, 155 
\bibitem[Strateva et al.(2001)]{2001AJ....122.1861S} Strateva, I., et al.\ 2001, AJ, 122, 1861 
\bibitem[Strauss et al.(2002)]{S02} Strauss, M.~A., et al.\ 2002, AJ, 124, 1810 
\bibitem[Teng et al.(2012)]{2012ApJ...753..165T} Teng, S.~H., Schawinski, K., Urry, C.~M., et al.\ 2012, ApJ, 753, 165 
\bibitem[Tojeiro et al. (2012)]{To12} Tojeiro, R., Masters, K.L, et al. \ 2012, MNRAS (submitted)
\bibitem[Trimble(1995)]{1995PASP..107.1133T} Trimble, V.\ 1995, PASP, 107, 1133 
\bibitem[van den Bergh(2011)]{2011AJ....141..188V} van den Bergh, S.\ 2011, AJ, 141, 188 
\bibitem[Villa-Vargas et al.(2010)]{2010ApJ...719.1470V} Villa-Vargas, J., Shlosman, I., \& Heller, C.\ 2010, ApJ, 719, 1470 
\bibitem[Wong et al.(2012)]{2012MNRAS.420.1684W} Wong, O.~I., Schawinski, K., Kaviraj, S., et al.\ 2012, MNRAS, 420, 1684 
\end{thebibliography}
\end{document}